# Lightweight 3D Convolutional Neural Network for Schizophrenia diagnosis using MRI Images and Ensemble Bagging Classifier

P SupriyaPatro[a], Tripti Goel[a,], S A VaraPrasad[a], M Tanveer[b], R Murugan[a]

[a] *Biomedical Imaging Lab, National Institute of Technology Silchar, Assam, 788010, Assam, India*
[b] *Department of Mathematics, Indian Institute of Technology, Indore, 453552, Madhya Pradesh, India*

**Abstract**

Structural alterations have been thoroughly investigated in the brain during the early onset of schizophrenia (SCZ) with the development of neuroimaging methods. The objective of the paper is an efficient classification of SCZ in 2 different classes: Cognitive Normal (CN), and SCZ using magnetic resonance imaging (MRI) images. This paper proposed a lightweight 3D convolutional neural network (CNN) based framework for SCZ diagnosis using MRI images. In the proposed model, lightweight 3D CNN is used to extract both spatial and spectral features simultaneously from 3D volume MRI scans, and classification is done using an ensemble bagging classifier. Ensemble bagging classifier contributes to preventing overfitting, reduces variance, and improves the model's accuracy. The proposed algorithm is tested on datasets taken from three benchmark databases available as open-source: MCICShare, COBRE, and fBRINPhase-II. These datasets have undergone preprocessing steps to register all the MRI images to the standard template and reduce the artifacts. The model achieves the highest accuracy 92.22%, sensitivity 94.44%, specificity 90%, precision 90.43%, recall 94.44%, F1-score 92.39% and G-mean 92.19% as compared to the current state-of-the-art techniques. The performance metrics evidenced the use of this model to assist the clinicians for automatic accurate diagnosis of SCZ.

*Keywords:* 3D-Convolutional Neural Network, Ensemble Bagging Classifier, Magnetic Resonance Imaging, Schizophrenia

# 1. Introduction

Schizophrenia (SCZ) is a chronic psychiatric disorder that generally appears in early adulthood or late teens and mainly influences how a person thinks, senses, and interacts. In the life of SCZ patients, 20%-40% of them have attempted suicide. Patients aged 25–34 years exhibit the highest suicide risk. Women showed a higher incidence of suicide by drowning and jumping than men [1]. SCZ is a heterogeneous disorder where the diagnosis is based on interviews and clinical symptoms. The symptoms of SCZ can be positive such as delusion, and hallucination, or can be negative symptoms such as abnormal motor behavior, disorganized thinking (speech), and poor social activities. The possible reasons for SCZ may be genetics, pregnancy or birth complications, previous drug use, and social factors [2]. Individuals with SCZ (IWSs) exhibit fewer overt expressions than non-patient comparison Subjects (NCSs) in verbal, facial, and acoustic channels. The three domains of emotion expression, emotion experience, and emotion recognition all exhibit dysfunction in IWSs [3], [4].

We can investigate structural brain abnormalities in SCZ patients like a decrease in brain volume, reduction in brain size, decrease in cortical thickness, and also some loss in grey matter with the help of Magnetic Resonance Imaging (MRI). MRI is a non-invasive and effective soft tissue contrast imaging modality [5] that offers information regarding the structure of tissue, such as its shape, size, and location, without exposing the matter to high ionization radiation. MRI research has produced more conclusive evidence of brain abnormalities in SCZ. The development of in vivoMRI techniques is largely responsible for this advancement in understanding the neuropathology of SCZ. As a result of these developments, a number of brain abnormalities in SCZ have now been discovered. Most of these abnormalities are small and subtle rather than large, necessitating the use of more sophisticated and precise measurement tools. Some of these abnormalities confirm earlier post-mortem findings. These findings include enlargement of the ventricles, and preferential involvement of the amygdala, hippocampus, parahippocampal gyrus, and temporal lobe regions of the neocortex. According to recent clinical findings, SCZ is characterized by deficits in cognitive function and brain structure, particularly inattention and decreased cortical thickness in related brain regions associated with increased peripheric inflammation [6].In this paper, we are concentrating on the cortical region because of its significance related to SCZ. Therefore, the structure of the brain should be investigated to make structural magnetic resonance imaging (sMRI) a potential tool for the efficient diagnosis of SCZ [7]. MRI images related to normal and SCZ are shown in Figure 1 (a) and (b).

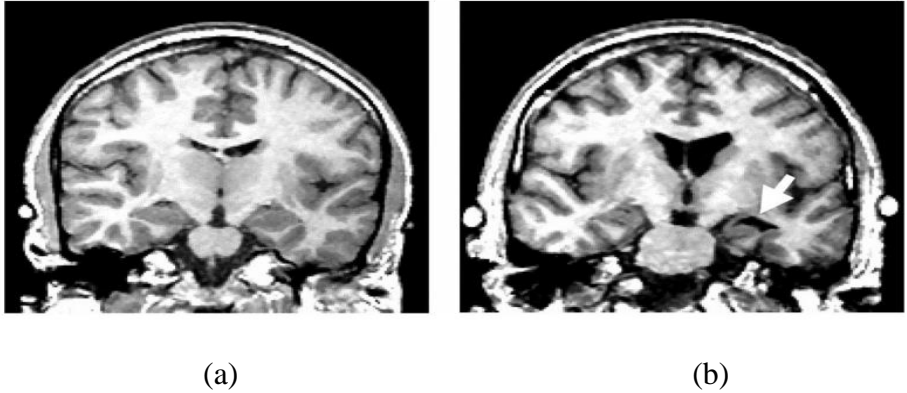

(a) (b)

Figure 1: (a)MRI Image for Normal (b)MRI Image for SCZ

Machine learning (ML) [8] is the field of study that gives the computer the capacity to learn without being explicitly programmed. Most common supervised algorithms for dealing with neuroimaging research include support vector machine (SVM) [9], decision tree (DT) [10], random forest (RF) [11], naive bayes (NB) [12], and artificial neural network (ANN) [13]. SVM finds an optimal boundary that maximizes the margin between support vectors and hyperplane [14]. The strength of the SVM depends on the kernel trick. Complex problems can also be solved by choosing the proper kernel function. But the selection of kernel functions is still a challenge that increases computational complexity. DT is structured with roots and edges. The most common limitations of DT are low prediction accuracy and a high probability of data overfitting. RF consists of multiple and diverse DTs together. RF is an integrated algorithm that gives better prediction performance and results than a single DT. The generalization error of a forest of tree predictors depends on an individual tree's strength and its correlation. NB is a simple and effective classifier algorithm that builds fast ML-based models by making quick predictions. NB classifier can handle both types of continuous and discrete data, are highly scalable, have a multi-class prediction, and requires less training data. However, NB assumes independent predictor features. ANN are widely used networks because of advantages like the ability to work with insufficient knowledge, parallel processing, requiring less statistical training, detecting non-linear complex relationships between dependent and independent variables, detect all possible interactions. However, with raw data, these ML algorithms do not work well and the features must be manually extracted. Because the selected features may not be trustworthy, handcrafted feature-based ML imposes limitations on the use of computer-aided diagnosis (CAD) in the actual world. ML-based on handcrafted features also lacks the capacity to extract additional features.

Deep learning (DL) [15,16] is a particular kind of ML algorithm inspired by the functionality of our brain cells called neurons which are implemented from the concept of ANN. There are several DL algorithms with different architectures which became popular due to their capabilities, accuracy, and efficiency in solving various problem domains. Convolutional neural networks (CNN) are a type of DL model that enhances the capacity of ANN by increasing the number of layers in the architecture. The advantages of CNN over other networks are, it can handle large and complex datasets, exploits spatial correlation, and employs multiple convolution layers. CNN automatically extracts the features which are most important for classification. CNN is classified as 2D or 3D CNN based on the type of input provided. 2D kernel predicts segmentation mapping for a single slice due to which they inherently fail to grip context from neighbor slices. 3D kernels address this issue by making segmentation maps for volumetric patches of a scan such as volume MRI data. Several applications based on shallow networks such as random vector functional link (RVFL) [17,18], and extreme machine learning (ELM)[19]also contribute to improved classification performance. A combination of deep shallow networks can provide much-improved performance by overcoming individual merits and demerits.

Ensemble models consist of several base learners to achieve better predictive performance than a single predictive model. Boosting, bagging, and stacking are the most popular ensemble methods. Boosting [20] is a technique that takes homogeneous weak learners, learns them sequentially in an adaptive manner, and then combines them to create a deterministic strategy. Initially, bootstrap samples acquired from the original training data were used to create an ensemble, thereafter each bootstrap sample was independently trained using base learners, and finally, the classification was done by voting [21]. Stacking takes into account non-homogeneous weak learners who learn them in parallel and incorporate them by training a meta-model to extract a prediction depending on the predictions of the various weak models. Ensemble bagging techniques are well suited for regression and classification. The novelty of the proposed algorithm is to preprocess the data in order to remove the artifacts and register the image with a standard MNI template. It will improve the generalization of the data such that it can be applied for the 3D CNN for SCZ diagnosis to a particular class. Then ensemble bagging is incorporated to classify the features which maintain the trade-off between overfitting and underfitting. These are the key contributions of the proposed paper.

- We have incorporated the preprocessing of data which will improve the learning efficiency of the model with the help of SPM12 software.

- 3D CNN is employed to extract complex features of high-dimensional images and extract the interslice information.

- Ensemble Bagging Classifier is used for the classification of the extracted features for improving the accuracy and robustness.

- We are comparing the proposed algorithm with other recent state of art classifiers to show the effectiveness of the proposed algorithm for SCZ diagnosis.

The whole paper is organized as follows: Section 2 discusses the literature review related to SCZ diagnosis. Section 3 introduces the data acquisition, preprocessing, and network architecture. Section 4 consists of implementation detail, results, and discussion. Finally, Section 5 will conclude the paper.

## 2. Literature Review

Some of the previous research done by the researcher for SCZ diagnosis have been discussed in this section. Chilla et al. [22] proposed the classification of SCZ and Healthy Control (HC) cohorts using a diverse set of neuroanatomical measures related to cortical and subcortical volumes and incorporated ensemble methods for better performance. Furthermore, they correlated such neuroanatomical features with quality of life (QoL) evaluation scores within the SCZ cohort. Classification accuracy results range from 83% to 87%, with sensitivity and specificity ranging from 90–98% and 65-70%, respectively. A robust interpretable (RobIn) deep network for diagnosing SCZ is proposed by Organisciak et al. [23] to minimize the training-application gap by concentrating on data that is easily available. Based on the DSM-5 criteria, gathered a data set of psychiatric observations of patients. Since all mental health clinics that diagnose SCZ using the DSM-5 already keep track of comparable data, the proposed method is easily incorporated into existing procedures as a tool to aid clinicians while adhering to formal diagnostic criteria. Subcortical regions and ventricles are identified as the most predictive brain regions through regional analysis. Human cognitive, affective, and social functions are largely mediated by subcortical structures, and structural abnormalities in these areas have been linked to SCZ. Zhang et al.[24] findings confirm that widespread changes in subcortical brain structureare linked to SCZ, and this structural information contributes significantly to diagnostic classification. Liu et al. [5] show the classification of SCZ patients and HC by extracting cortical thickness totally independent at various regions of interest based on a hierarchical network and then SCZ is classified using two-step feature selection to acquire two optimal kernel ma- trices using SVM radial basis function (SVM-RFB). Finally, multiple kernel learning (MKL) is used to generate a combination of kernels and learn based on optimal kernel matrices.

Tandon et al. [25] proposed the techniques of ML and their application to study the SCZ. Vyskovsky et al. [26] provide two automated whole-brain morphometry methods: Voxel-Based Morphometry (VBM) and Deformation Based Morphometry (DBM) for SCZ diagnostics. In this paper multilayer perceptron (MLP) is used as a classifier. The features are extracted from VBM, DBM individually, and a combination of both VBM and DBM. The results show a 5% increment in accuracy for MLP classification based on combining VBM and DBM models than individual MLP classification on VBM or DBM models. Structural MRI data with nonlinear independent component estimation (NICE) is proposed to discover aberrant samples of grey matter concentration in SCZ patients. For this biomedical application, Castro et al.[27] deal with the issue of nonlinear ICA model regularization by employing dimensionality reduction, as well as adequate control of the model's complexity and the use of a suitable approximation of the probability distribution function of the components estimated.

Lin et al. [19] apply an Ensemble Bagging framework with a feature-extracting algorithm. This feature selection is based on the analysis of three clinical symptom scales and eleven cognitive factors. The author of this paper proposes a connection between cognitive function and functional outcome in SCZ. It is clear from the analysis of the results that Ensemble Bagging with feature extraction outperforms another model. Ventura et al. [28] provide a meta-analysis that is guided to determine the magnitude of the connection between cognitive function and functional outcome in SCZ. In this paper, a model was generated to test which symptom is more connected to neurocognition and functional outcome. Results show that negative symptoms were found to have a significant relationship with neurocognitive functioning, whereas positive symptoms were not.

Goceri et al. [29] propose Sobolev gradient-based optimization (SGB) and 3D CNN for the diagnosis of AD. In this paper, a robust and deeply supervised 3D CNN model with max-pooling function and leaky ReLU is developed using 3D features to improve the accuracy of MRI scans. The 3D CNN model is integrated with the new SGB optimizer. The proposed model with automated diagnosis extracts desired features with the best accuracy. Hu et al. [30] proposed a deep feature that depends on pre-trained 2D CNN and naive 3D CNN models for SCZ classification by incorporating 3D structural with diffusion MRI data and revealed that 3D CNN performs better than the pretrained 2D CNN model. Furthermore, identifies the grey matter and white matter regions of the brain that are affected. Oh et al. [31] proposed 3D Convolutional Auto Encoder (CAE) CNN model. In this paper, the 3D activational map is used as input to preserve the spatial locality.

The performance of 3D CNN models with various architectures is compared by Hu et al.[32]. CNN models outperformed manually created feature-based machines in terms of cross-validation accuracy. The sample size for this study is relatively small, particularly for network training, which may have an impact on the model performance and result in lower accuracy and generalizability. Qureshi et al.[33] adopted a 3D-CNN DL architecture by Independent Component Analysis (ICA). In this study, the feature selection method was not employed. Therefore, it is difficult to determine which ICA components contributed more to the classification. By combining 3D structural and diffusion MRI data, M. Hu et al.[34] developed a multimodal 3D CNN model with various architectures and used a deep feature approach based on 2D pre-trained CNN. In order to lower the GPU memory requirement, the input maps are downscaled in this study. The GPU memory limitations also have an impact on the model depths. The above papers give the motivation to address the issues we proposed a Lightweight 3D CNN for SCZ diagnosis using Ensemble Bagging Classifier. The proposed model achieves better accuracy compared to SVM and other 3D CNN models. After a review of the above literature work, it is evident that CNN can generate accurate results and is good for feature extraction.

The primary goal of the present study is to explore whether the proposed model using Ensemble Bagging as a classifier produces a better accuracy compared to traditional methods. We have proposed an efficient framework that is employed with the integration of lightweight 3D CNN and Ensemble Classifier to classify the particular class. In this architecture 3D, CNN is used for feature detection to extract both spatial and spectral features simultaneously from 3D MRI scans. Ensemble Bagging is used as a classifier to minimize variance and enhance the performance of the model.

## 3. Methodology

A detailed explanation of the proposed architecture is represented in the below subsections. The flowchart of the proposed architecture is shown in Figure 2.

*Data acquisition*

The structural magnetic resonance imaging (sMRI) images of CN and SCZ patients are collected from the publicly available dataset: MCICShare, COBRE, and fBRINPhase-II. The initial sub-dataset is MCICShare, which consists of 95 SCZ patients and 45 CN were obtained with the following parameters: repetition time (TR)=2000ms, echo time (TE)= 30ms, flip angle of 9 degrees and slice thickness is 4mm. The second sub-dataset is COBRE, which consists of 84 SCZ patients and 34 CN were obtained with the following parameters: TR=2000ms, TE= 29ms, flip angle of 7 degrees and slice thickness is 1mm. The final sub-dataset is fBRINPhase-II, which consists of 86 SCZ patients and 30 CN were obtained with the respective parameters: TR=2500ms, TE= 27ms, flip angle of 7 degrees and slice thickness is 1mm. The entire neuroimage scans are T1- weighted MRI scans. A total of 600 participants undergone for sMRI are taken of which, 300 subjects are CN and 300 subjects are SCZ patients.

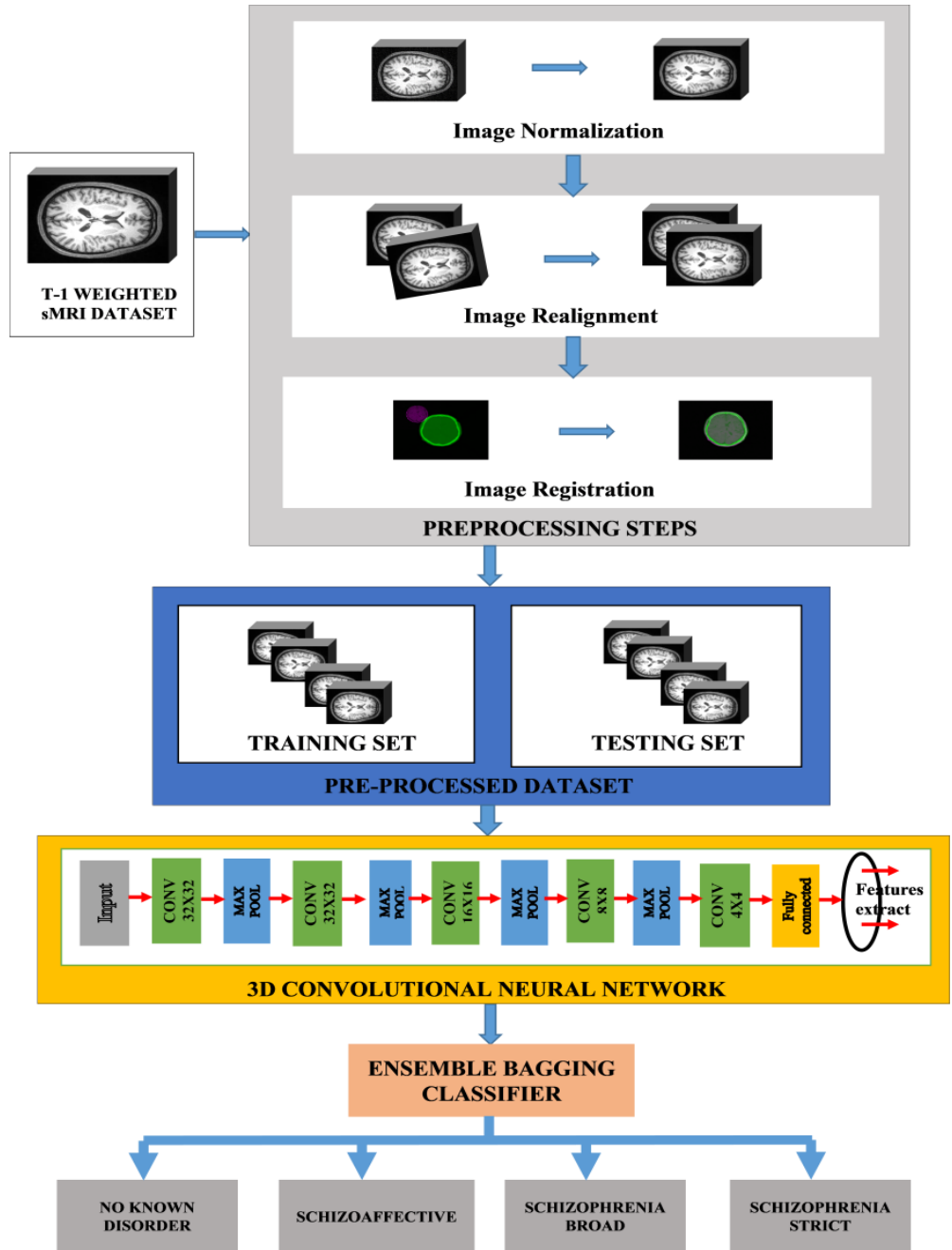
Figure 2: The Proposed architecture of SCZ diagnosis

*Preprocessing steps*

As data is acquired from different sources, preprocessing is done to normalize, realign and register all the MRI scans. A detailed description of preprocessing steps is presented in this subsection. All the preprocessing steps are carried out by SPM12.

The initial step in preprocessing is normalization. The purpose of normalization is to standardize all of the input images in a template. This MRI scan's intensity is normalized by subtracting the mean intensity value and then dividing the variance, resulting in a scan with a zero mean and unit variance. Figure 3 (a) and (b) show the results before and after normalization, respectively.

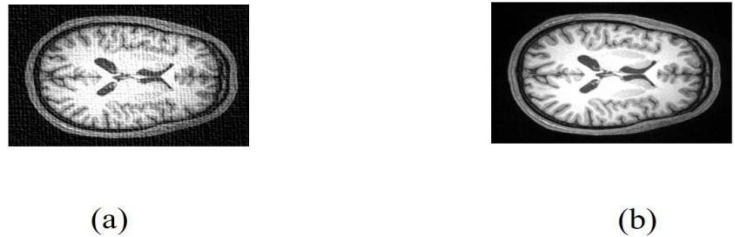

(a) (b)

Figure 3: (a) Image before normalization and noise removal (b) Image after normalization and noise removal

The second step in preprocessing is image realignment, which has the primary goal of removing artifacts caused by the patient's motion during the MRI scan. The scan is manipulated using rigid body transformation. Estimation and reslicing are the most commonly used methods for realignment. In estimation and reslicing, interpolation is a method using by which images are sampled while estimating the optimum transformation. Interpolation is better when the order of interpolation is higher. So, the 4th order B-spline is done on the scans and no wrapping is done. The results of the image before and after realignment is shown in Figure 4 (a) and (b) respectively.

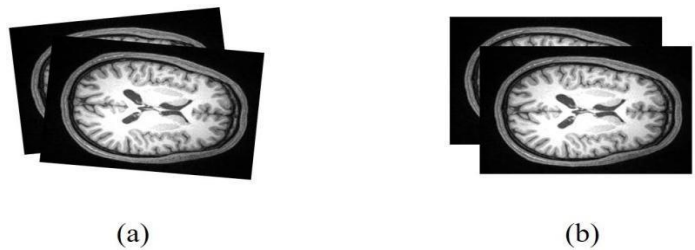

(a) (b)

Figure 4: (a) Image before Realignment (b) Image after Realignment

Finally, image registration is performed by applying the geometric transformation which can be parameterized by three translations and three rotations about the different axes. Every scanner has different scanning parameters such as the number of slices, slice thickness, flip angle, etc. So, to make every image match the template each brain needs to be transformed to have the same shape, size, and dimensions. In Image registration, estimation and reslicing are done by taking the Average distance between the sample points as 4mm*2mm, and Gaussian histogram smoothing of 7*7 is done in order to make the cost function as smoothly as possible, to increase the rate of convergence and less chance of local minima. The image before and after registration is shown in Figures 5 (a) and (b) respectively.

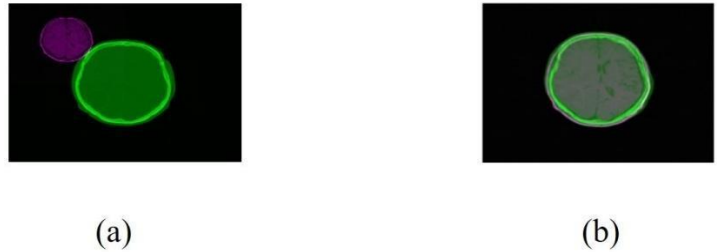

Figure 5: (a) Image before Registration (b) Image after Registration

After preprocessing, the slices have been extracted in order to perform a comparison between the 2D and 3D approaches. A total 1500 slice has been extracted from 300 subjects in order to perform classification. MRICRON has been used for extracting the slices from 3D NIfTI.

*Network architecture*

CNN is one of the most successful approaches for feature extraction and classification of data such as video, image, and speech. CNN consists of 3 main layers: Input, hidden, and output. This hidden layer is further split into the convolution layer (CL), rectified linear unit (ReLU), and max-pooling layer (MPL). The CL is the first layer of 3D CNN architecture which is employed to extract features from the input layer. Continual application of the same filter over the input layer results in a map of activation called a feature map, which states the location and robustness of the detected features from the input image. This obtained feature is passed through a nonlinear activation function such as ReLU for non-linear mapping with feature space. A batch normalization layer is used between CL and ReLU layers. The MPL is employed to reduce computational power by minimizing the dimension of the feature map and preserving the most relevant information about the image. The labeled model of lightweight 3D CNN is illustrated in Figure 6.

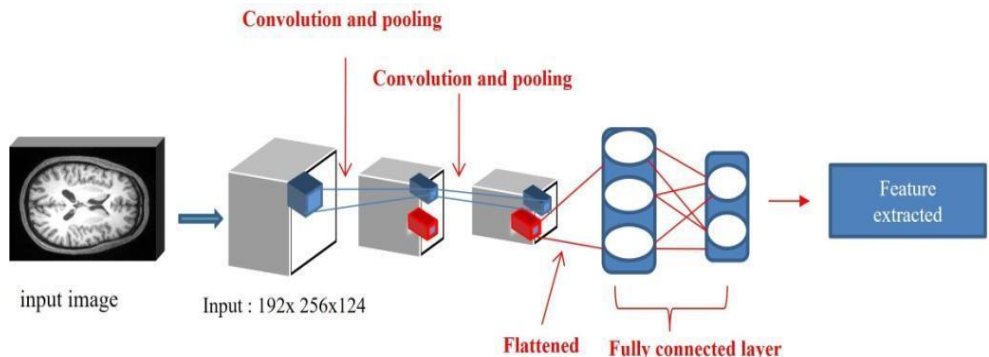

Figure 6: 3D-convolutional neural network

The features extracted from the fully connected layer (FCL) are given to the Ensemble Bagging Classifier to classify the input features into a particular class. A bagging classifier is an ensemble technique proposed by Leo Breiman [35]. The basic idea of the Ensemble Bagging Classifier is explained in Figure 7.

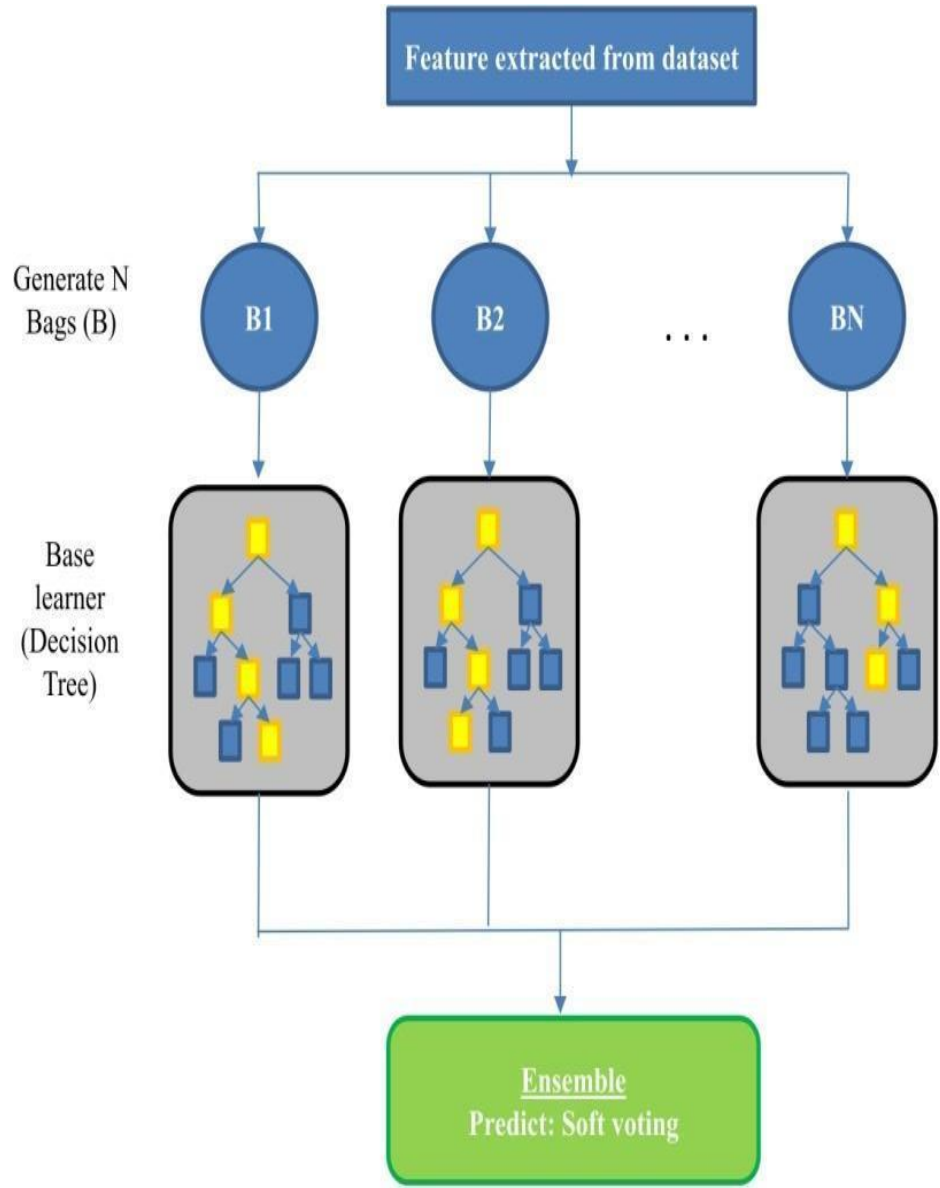

Figure 7: Ensemble bagging classifier

The main aim of the Ensemble technique is to achieve low bias and low variance by combining multiple base learners. The base learners are the independent models that combine during ensembling to achieve the final output.

In the proposed paper, the base learner model is considered to be DT. In Ensemble Bagging Classifier DT as the base learner is applied for each N partition (or bags) by bootstrap sampling and use one partition for each base learner which is generated using the DT algorithm. This base learner gets trained by their respective data bag and after generating the base learner, it goes into a parallel process for generating the next base learner. The final estimator is done by voting for the classification from which the ensemble prediction is done. In this proposed work, soft voting is used to calculate probabilities of the outcomes of classifiers. This soft voting gives the best result by averaging probabilities of the outcomes of the individual algorithm.

## 4. Results and Discussion

*Implementation detail*

In this paper, the proposed model is implemented using MATLAB 2021a and executed on Windows 10 with an Intel(R) Xenon(R) W-2133 CPU @3.60 GHz with 64 GB RAM. The GPU used is NVIDIA QuadroP2200. The proposed model is tested for CN and SCZ patients using a publicly available dataset. The ratio of training and testing is 70:30, respectively.

*Performance metrics*

The proposed model's main aim is to diagnose SCZ from MRI images automatically. To analyze the performance of the proposed model, the proposed algorithm's accuracy, sensitivity, specificity, precision, recall, F-measure, and G-mean are contrasted with state-of-the-art methods. Accuracy measures the overall system performance and the ability of the classifier to differentiate between SCZ and CN. Sensitivity evaluates the ratio of true positives to the total positives in the CM, while specificity evaluates the ability of the model to predict true negatives. Precision evaluates the number of correct positive predictions done among all positive predictions, whereas Recall indicates missed positive prediction. F-measure provides the model's accuracy on the dataset. G-mean evaluates the balance between the classifications performances on different classes available, lower G-mean indicates poor performances of the model to classify the positive class.

The ROC and CM is the pictorial representation for measuring the effectiveness of classification. CM analysis is deeper than classification accuracy by displaying true or false predictions for each class. The ROC curves show the classifier's performance in graphical form at all the classification thresholds. This curve plot the parameters on account of the true-positive rate (TPR) and false-positive rate (FPR).

*Comparison with state-of-the-art classifier*

An analysis based on performance is done with a state-of-the-art classifier, as shown in Table 1, and CM and ROC curves are obtained as shown in Figures 8 and 9 respectively. The performance matrices of the Ensemble Bagging Classifier are compared with other state-of-art based on accuracy, sensitivity, specificity, precision, recall, f-measure, and G-mean. The acquired performance parameter for the Ensemble Bagging Classifier is found to be better than other state-of-the-art classifiers.

Table 1: Comparison of Ensemble Bagging Classifier with state-of-the-art classifier

| Classifier | Accuracy | Sensitivity | Specificity | Precision | Recall | F-measure | G-mean |
|---|---|---|---|---|---|---|---|
| SVM | 91.67 | 90 | 93.33 | 93.10 | 90 | 91.53 | 91.64 |
| Naive Bayes | 90.56 | 90 | 91.11 | 91.01 | 90 | 90.50 | 90.55 |
| K-Nearest Neighbour | 90 | 86.67 | 93.33 | 92.86 | 86.67 | 89.66 | 89.93 |
| Random Forest | 91.11 | 90 | 92.22 | 92.05 | 90 | 91.01 | 91.10 |
| Standard RVFL | 91.66 | 90 | 93.33 | 93.10 | 90 | 91.52 | 91.64 |
| **Ensemble Bagging** | **92.22** | **94.44** | **90** | **90.43** | **94.44** | **85.71** | **93.40** |

However, RVFL and SVM classifiers specificity obtained is better than the proposed approach, but the classification accuracy was higher in all case for the proposed model. The advantage of using Ensemble Bagging as a classifier is that it reduces variance and maintains bias. Thus, the Ensemble Bagging technique as a classifier can acquire high accuracy compared to other classification techniques.

*Comparison with different Lightweight CNN architecture*

The performance analysis of different architectures is shown in Table, and the developed architecture is shown in Figure 10. Further, the performance is analyzed by obtaining CM and ROC curves, as shown in Figures 10 and 11 respectively. An experiment is conducted to compare the proposed architecture (Network 1) with different 3D CNN architecture (Network 2, Network 3, Network 4, and Network 5) by increasing or decreasing the number of convolution layers. The proposed architecture gives better accuracy compared to other lightweight 3D CNN architecture.

Table 2: Performance analysis of different architecture

| Architecture | Accuracy | Sensitivity | Specificity | Precision | Recall | F-measure | G-mean |
|---|---|---|---|---|---|---|---|
| **Network 1-3D** | **92.22** | **94.44** | **90** | **90.43** | **94.44** | **85.71** | **93.40** |
| Network 1-2D | 90.17 | 90.67 | 89.67 | 89.77 | 90.67 | 90.22 | 90.17 |
| Network 2-3D | 88.33 | 83.33 | 93.33 | 92.59 | 83.33 | 87.72 | 88.19 |
| Network 2-2D | 90.50 | 90.33 | 90.67 | 90.64 | 90.33 | 90.48 | 90.50 |
| Network 3-3D | 80 | 78.88 | 81.11 | 80.68 | 78.88 | 79.76 | 79.98 |
| Network 3-2D | 87.67 | 91 | 84.33 | 85.31 | 91 | 88.06 | 87.60 |
| Network 4-3D | 88.89 | 87.78 | 90 | 89.77 | 87.78 | 88.76 | 88.88 |
| Network 4-2D | 91.83 | 90.67 | 93 | 92.83 | 90.67 | 91.74 | 91.83 |
| Network 5-3D | 86.11 | 97.78 | 74.44 | 79.28 | 97.78 | 87.56 | 85.32 |
| Network 5-2D | 89.33 | 92.67 | 86 | 86.88 | 92.67 | 89.68 | 89.27 |

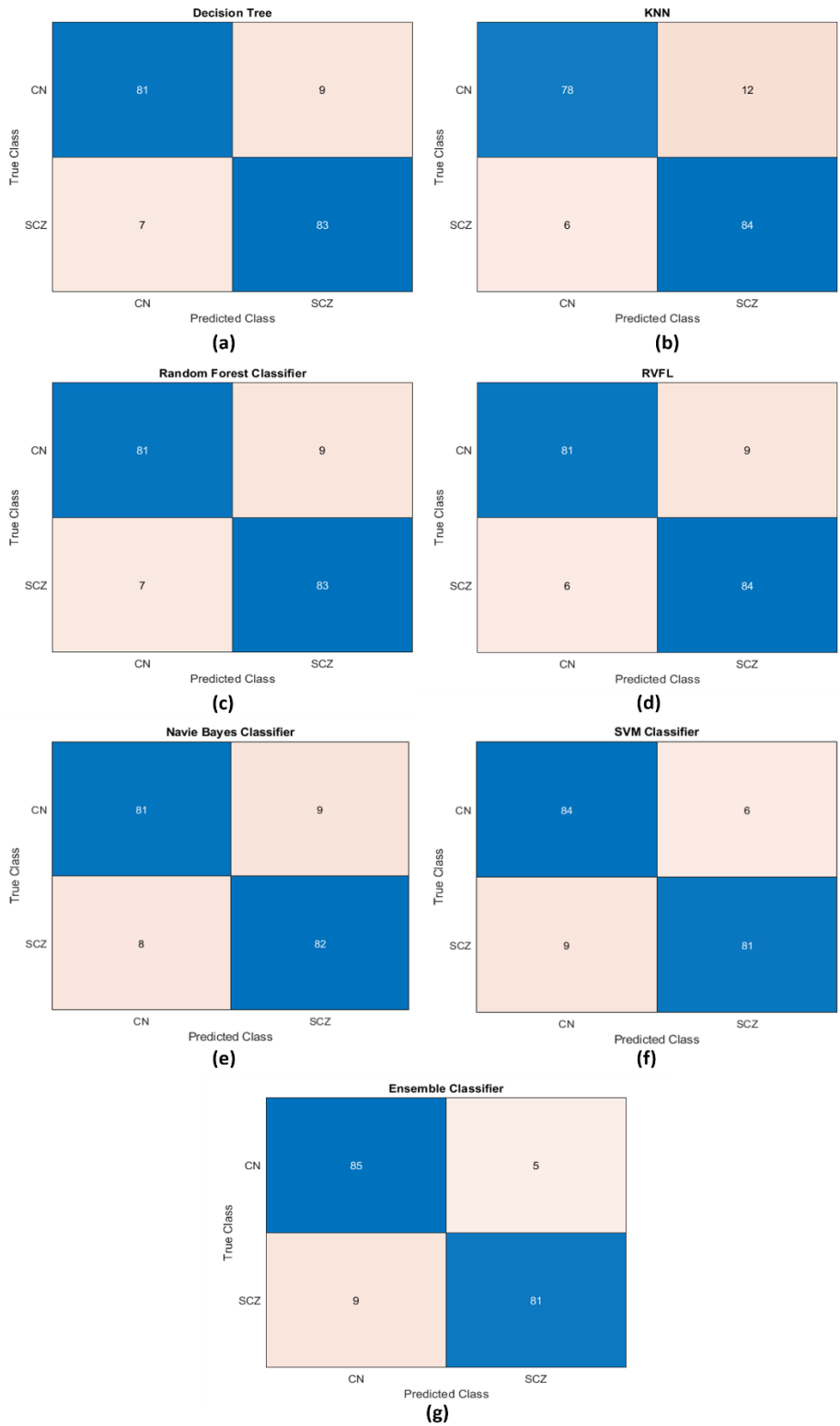

Figure 8: Confusion matrix of the (a) DT (b) K-Nearest Neighbour (c) Random forest (d) RVFL (e) Naive bayes (f) SVM and (g) Ensemble bagging classifier

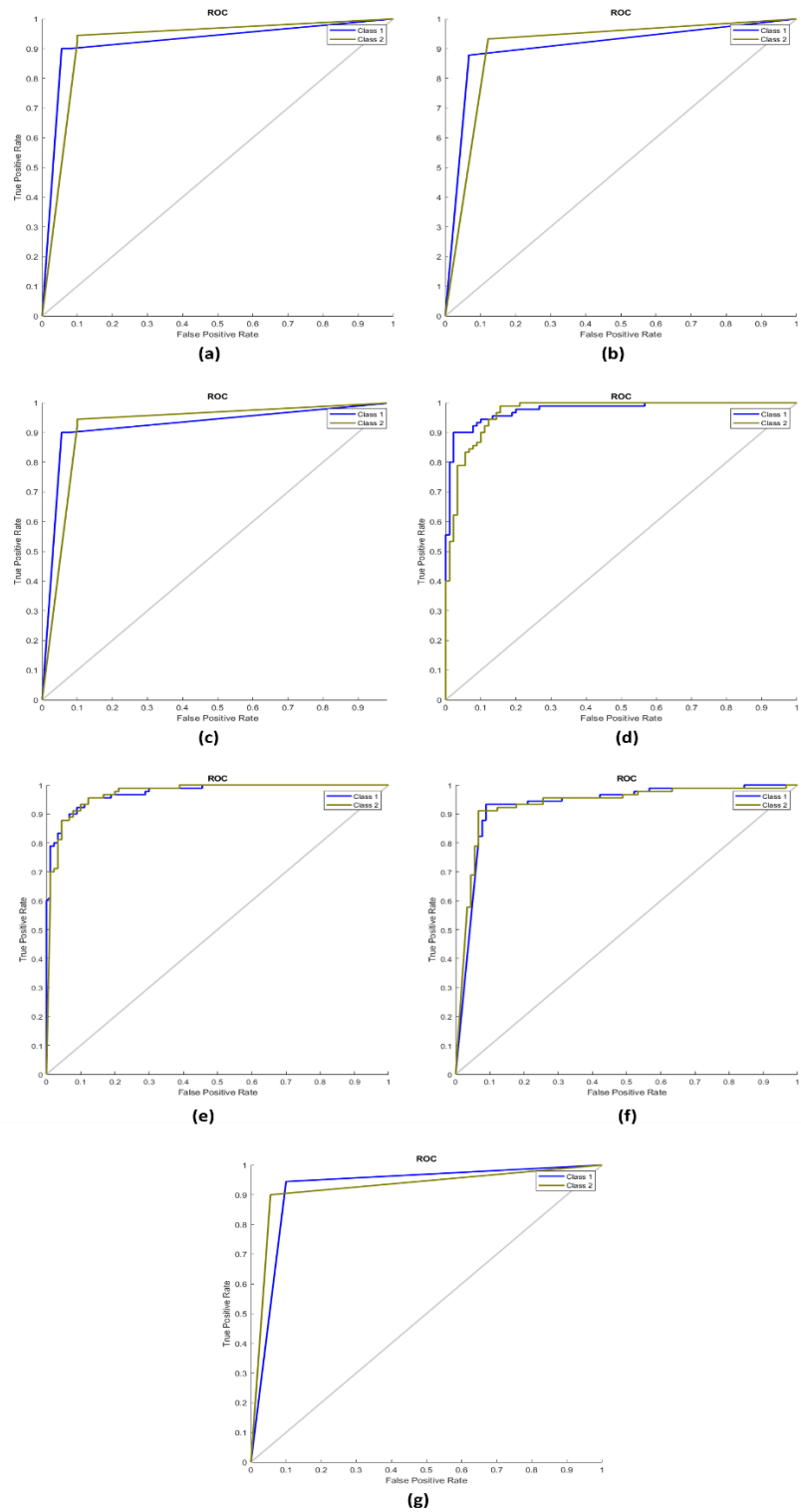

Figure 9: ROC of the (a) DT (b) K-Nearest Neighbour (c) Random forest (d) RVFL (e) Naive bayes (f) SVM and (g) Ensemble bagging classifier (Class 1-CN, Class 2-SCZ)

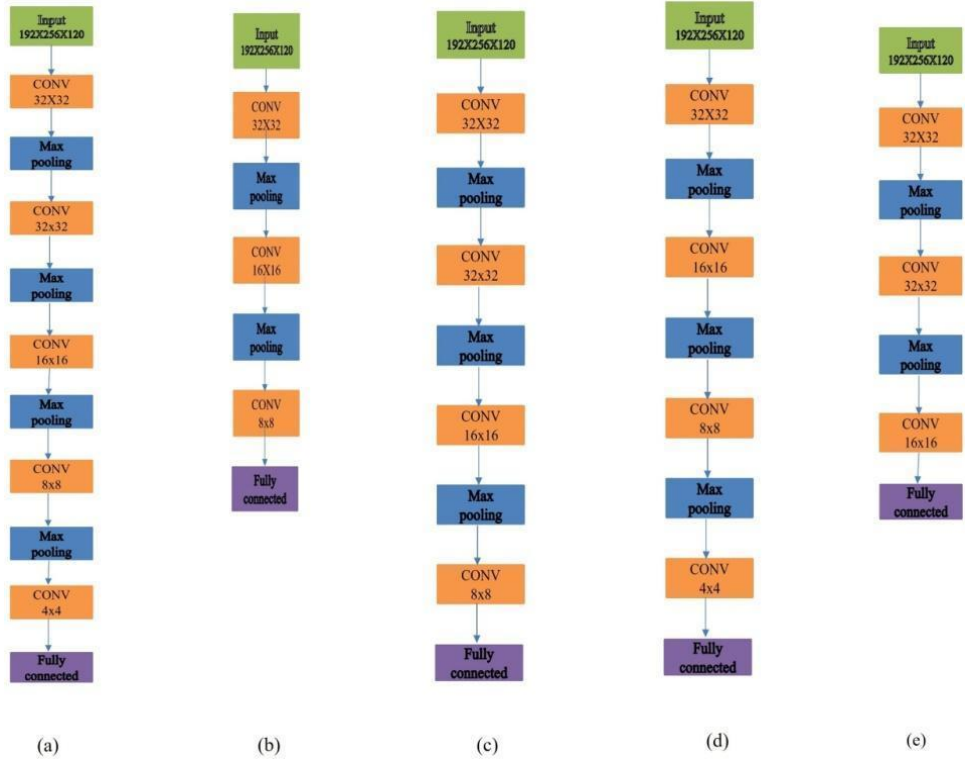

Figure 10: Different architecture of CNN (a)Network 1 (b)Network 2 (c)Network 3 (d)Network 4 and (e) Network 5

*Computational Complexity*

The computational complexity of 3D CNN network is calculated as $O(MNKn^2 ctW)$. In this equation, $M N and K$ are the dimension of the 3D image, $n \times n$ is the size of the filter. c is the number of channels and t is the temporal dimension and W is the number of filters.

The computational complexity of the ensemble bagging classifier is $O(BD)$, where B is the number of the bags and D is the number of classifiers. Therefore, the total computational complexity of the proposed network is $O(MNKn^2 ctW) + O(BD)$.

*Discussion*

The main objective of this paper is to accurately detect SCZ so that proper care and treatment for the patient can be taken which can help to reduce the mortality ratio. We have to start with preprocessing of data in order to get better visualization and improve the learning efficiency of the model. we incorporate the 3D CNN for extracting voxel information. Finally, Ensemble Bagging as a classifier for classification.

Table 2 shows the results of the comparison of the proposed novel architecture with state-of-the-art classifiers. As seen from the table Ensemble Bagging Classifier outperforms state-of-the-art classifiers. The results of the KNN classifier are relatively close to the Ensemble Bagging Classifier in terms of specificity.

Table 3 shows the Performance analysis of the different architectures of CNN. According to the results, Network 1 with 5 layered CNN shows superior performance than other networks in terms of Accuracy, Sensitivity, Recall, F-measure, and G-mean. Therefore, the proposed 3D model can differentiate SCZ patients from CN with much accuracy and more feature learning.

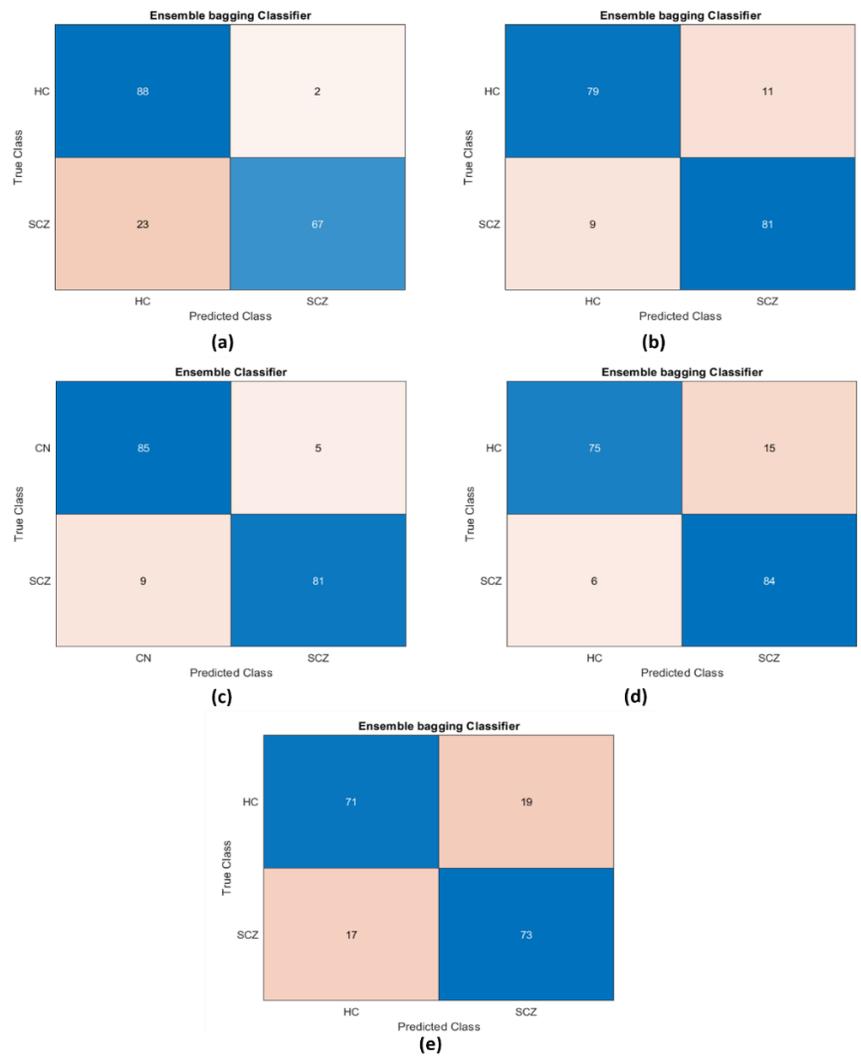

Figure 11: Confusion matrix of the different 3D (a)Network 1 (b)Network 2 (c)Network 3 (d)Network 4 and (e) Network 5

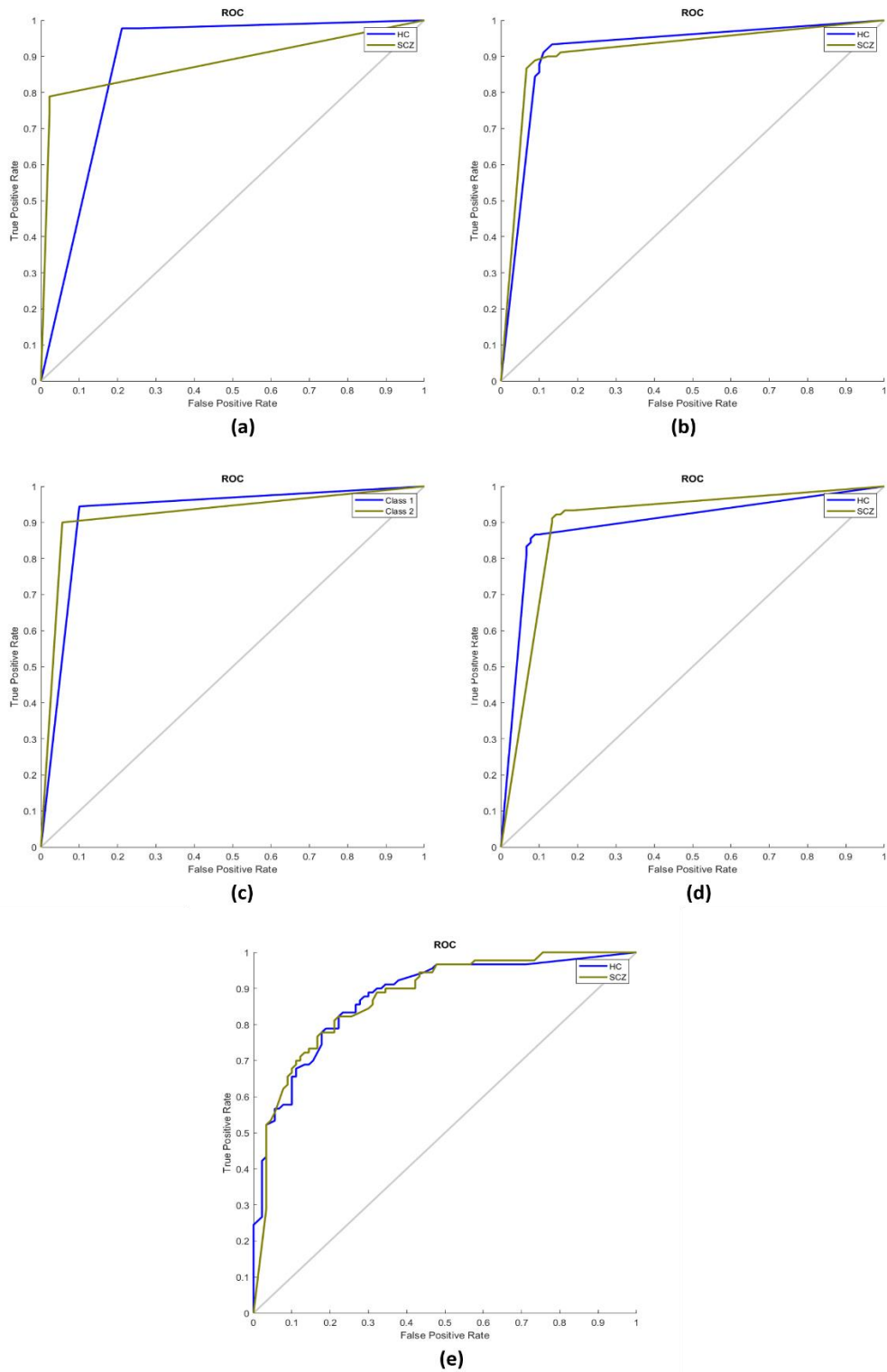

Figure 12: ROC curve of the 3D network with different layers (a) Network 1 (b) Network 2 (c) Network 3 (d) Network 4 and (e) Network 5 (Class 1-CN, Class 2-SCZ)

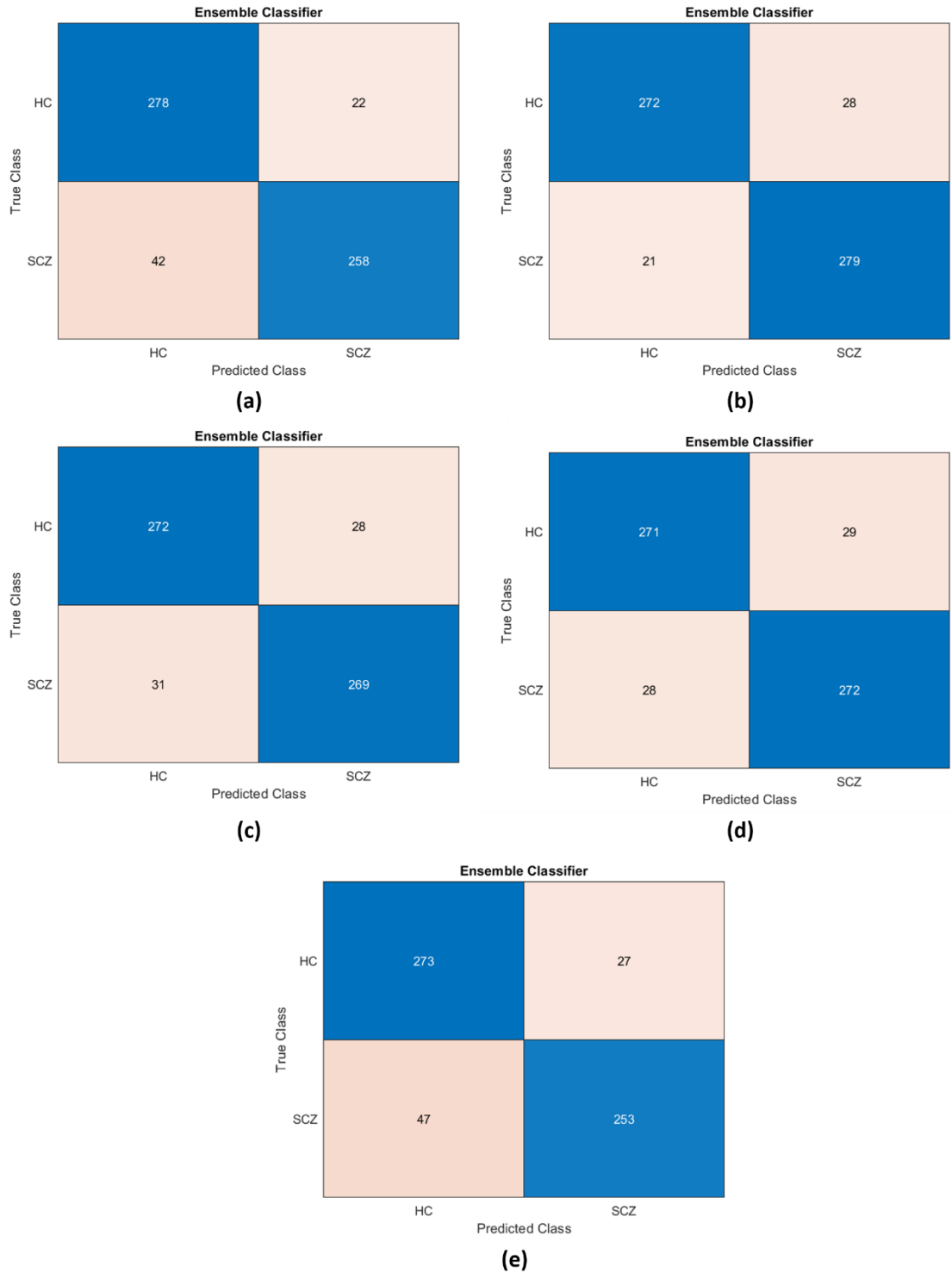

Figure 13: Confusion matrix of the different 2D network with different layers (a)Network 1 (b)Network 2 (c)Network 3 (d)Network4 and (e) Network 5

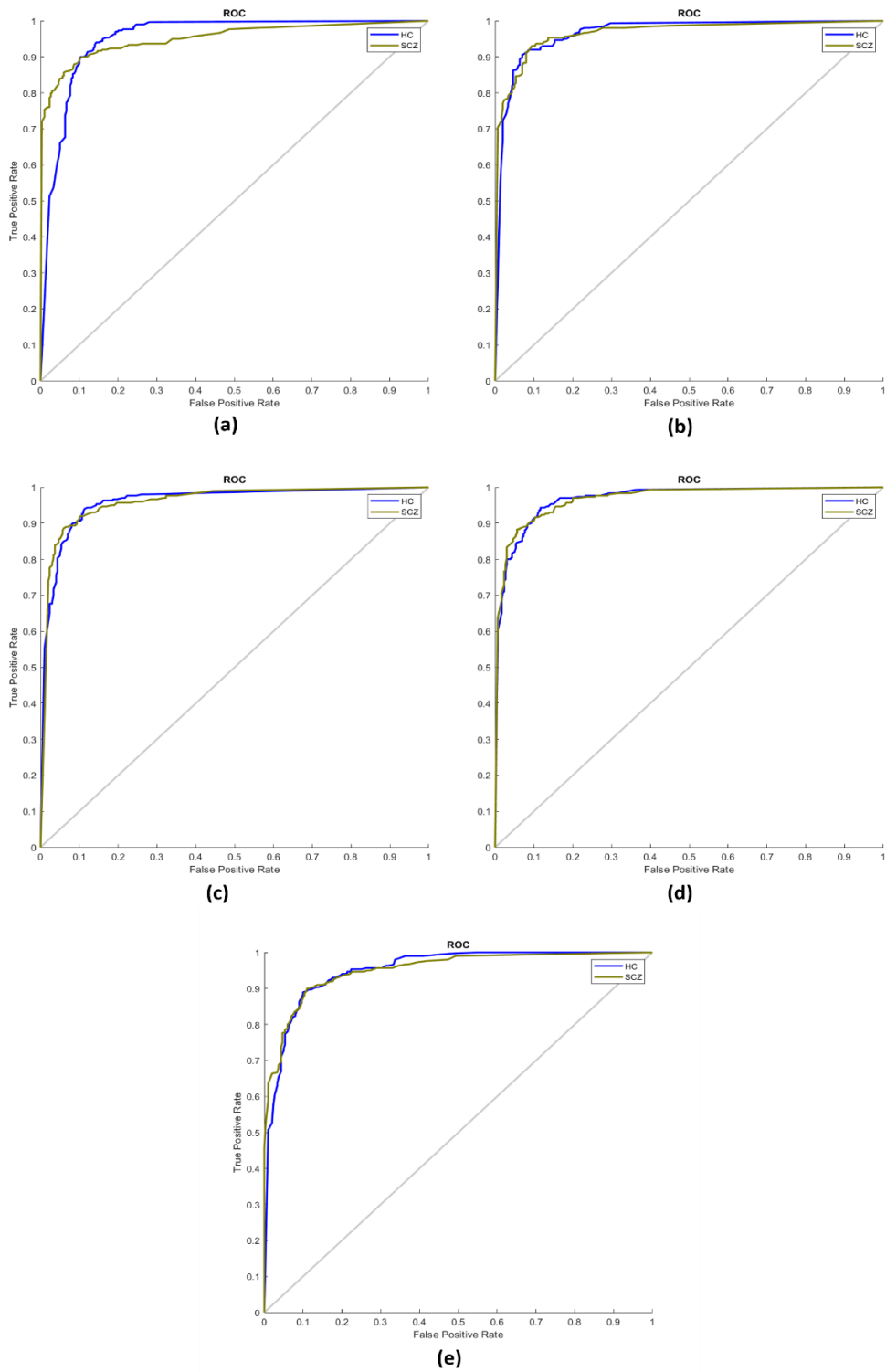

Figure 14: ROC curve of the 2D network with different layers (a) Network 1 (b)Network 2 (c) Network 3 (d) Network 4 and (e) Network 5 (Class 1-CN, Class 2-SCZ)

So, the care and treatment are done according to the severity of the disease. we can conclude that the proposed network performs better than another state-of-the-art. The proposed model's CNN architecture is compared with other CNN architecture for further analysis of its performance, and from the outcome of the results, we can conclude that the proposed model achieves the highest accuracy of 92.22%. Therefore, this model can be used to assist clinicians in the automated diagnosis of SCZ.

## 5. Conclusion and Future Scope

This paper represents a novel architecture of integration of lightweight 3D CNN with an Ensemble classifier for the diagnosis of SCZ. The features of sMRI images are extracted using 3D CNN. We employed a 3D CNN for feature extraction because of the ability to leverage interslice context. In order to maintain bias and variance tradeoffs and to prevent overfitting, Ensemble Bagging is integrated with 3D CNN as a classifier. As a result, the novel architecture model results in improved performance and high accuracy. The classification evaluations also revealed that the optimal feature set processed through 3D CNN can significantly discriminate between normal and SCZ images and shows an effective performance compared to previous classification methods. There are certain limitations of the proposed model. Firstly, training the 3D models increases the computational cost, but the 3D model extracts high dimensional voxel information compared to 2D models. Second, the binary labels in the MRI data are used to train the DL algorithm (No known disorder, Schizoaffective, Schizophrenia broad, Schizophrenia). The majority of psychiatric disorders progress along a continuum, and a patient may have multiple conditions at once. Further research involving other psychiatric conditions, such as bipolar spectrum and neurodegenerative disorders, is required because our analysis lacked a clinical comparison group. Due to the absence of clinical control groups, it is challenging to determine whether the characteristics of SCZ observed are unique to SCZ. Finally, DT is used as a base learner, the accuracy of DT can get changed depending on the depth of the DT and the continuous split of nodes of the tree may lead to overfitting.

In the future, the proposed algorithm can be implemented for any other psychotic disorder. The base learner other than DT can be used to further improve accuracy. Further, this proposed network can be applied to a functional MRI dataset to diagnose the SCZ disease accurately.

## References

[1] C.-H. Pan, P.-H. Chen, H.-M. Chang, I. Wang, Y.-L. Chen, S.-S. Su, S.Y. Tsai, C.-C. Chen, C.-J. Kuo, et al., Incidence and method of suicide mortality in patients with schizophrenia: a nationwide cohort study, Social psychiatry and psychiatric

epidemiology 56 (8) (2021) 1437–1446.

[2] X. Zhang, Four possible causes of schizophrenia, in: 2020 International Conference on Public Health and Data Science (ICPHDS), IEEE, 2020, pp. 320–323.

[3] F. Tr´emeau, A review of emotion deficits in schizophrenia, Dialogues in clinical neuroscience (2022).

[4] Tanveer, Muhammad, Jatin Jangir, M. A. Ganaie, Iman Beheshti, M. Tabish, and Nikunj Chhabra. "Diagnosis of Schizophrenia: A comprehensive evaluation." IEEE Journal of Biomedical and Health Informatics (2022).

[5] J. Liu, M. Li, Y. Pan, F.-X. Wu, X. Chen, J. Wang, Classification of schizophrenia based on individual hierarchical brain networks con- structed from structural MRI images, IEEE transactions on nanobioscience 16 (7) (2017) 600–608.

[6] H. F. North, J. Bruggemann, V. Cropley, V. Swaminathan, S. Sun- dram, R. Lenroot, A. M. Pereira, A. Zalesky, C. Bousman, C. Pantelis, Increased peripheral inflammation in schizophrenia is associated with worse cognitive performance and related cortical thickness reductions, European Archives of Psychiatry and Clinical Neuroscience 271 (4) (2021) 595–607.

[7] J. Kambeitz, L. Kambeitz-Ilankovic, S. Leucht, S. Wood, C. Davatzikos, B. Malchow, P. Falkai, N. Koutsouleris, Detecting neuroimaging biomarkers for schizophrenia: a meta-analysis of multivariate pattern recognition studies, Neuropsychopharmacology 40 (7) (2015) 1742–1751.

[8] S. Lemm, B. Blankertz, T. Dickhaus, K.-R. Mu¨ller, Introduction to machine learning for brain imaging, Neuroimage 56 (2) (2011) 387–399.

[9] C. J. Burges, A tutorial on support vector machines for pattern recognition, Data mining and knowledge discovery 2 (2) (1998) 121–167.

[10] J. Diederich, A. Al-Ajmi, P. Yellowlees, Ex-ray: Data mining and mental health, Applied Soft Computing 7 (3) (2007) 923–928.

[11] L. Breiman, Random forests, Machine learning 45 (1) (2001) 5–32.

[12] N. Friedman, D. Geiger, M. Goldszmidt, Bayesian network classifiers, Machine learning 29 (2) (1997) 131–163.

[13] G. R. Yang, X.-J. Wang, Artificial neural networks for neuroscientists: A primer, Neuron 107 (6) (2020) 1048–1070.

[14] C. Cortes, V. Vapnik, Support-vector networks, Machine learning 20 (3) (1995) 273–297.

[15] J. A. Cortes-Briones, N. I. Tapia-Rivas, D. C. D'Souza, P. A. Estevez, Going deep into schizophrenia with artificial intelligence, Schizophrenia Research 245 (2022) 122–140.

[16] Sharma, Rahul, Tripti Goel, Muhammad Tanveer, and R. Murugan. "FDN-ADNet: Fuzzy LS-TWSVM based deep learning network for prognosis of the Alzheimer's disease using the sagittal plane of MRI scans." Applied Soft Computing 115 (2022): 108099.

[17] Shi, Qiushi, Rakesh Katuwal, Ponnuthurai N. Suganthan, and Muhammad Tanveer. "Random vector functional link neural network based ensemble deep learning." Pattern Recognition 117 (2021): 107978.

[18] Malik, Ashwani Kumar, M. A. Ganaie, M. Tanveer, and P. N. Suganthan. "Extended features based random vector functional link network for classification problem." IEEE Transactions on Computational Social Systems (2022).

[19] Huang, Gao, Guang-Bin Huang, Shiji Song, and Keyou You. "Trends in extreme learning machines: A review." Neural Networks 61 (2015): 32-48.

[20] E. Lin, C.-H. Lin, H.-Y. Lane, Applying a bagging ensemble machine learning approach to predict functional outcome of schizophrenia with clinical symptoms and cognitive functions, Scientific Reports 11 (1) (2021) 1–9.

[21] S. Sreng, N. Maneerat, K. Hamamoto, R. Panjaphongse, Automated diabetic retinopathy screening system using hybrid simulated annealing and ensemble bagging classifier, Applied Sciences 8 (7) (2018) 1198.

[22] G. S. Chilla, L. Y. Yeow, Q. H. Chew, K. Sim, K. Prakash, Machine learning classification of schizophrenia patients and healthy controls using diverse neuroanatomical markers and ensemble methods, Scientific reports 12 (1) (2022) 1–11.

[23] D. Organisciak, H. P. Shum, E. Nwoye, W. L. Woo, Robin: A robust interpretable deep network for schizophrenia diagnosis, Expert Systems with

Applications 201 (2022) 117158.

[24] J. Guo, J. Zhang, V. Rao, Y. Tian, Y. Yang, N. Acosta, Z. Wan, P.Y. Lee, C. Zhang, L. Kegeles, et al., Detecting schizophrenia with 3d structural brain mri using deep learning (2022).

[25] N. Tandon, R. Tandon, Using machine learning to explain the heterogeneity of schizophrenia. realizing the promise and avoiding the hype, Schizophrenia Research 214 (2019) 70–75.

[26] R. Vyˇskovsky`, D. Schwarz, T. Kaˇspˊarek, Brain morphometry methods for feature extraction in random subspace ensemble neural network classification of first-episode schizophrenia, Neural computation 31 (5) (2019) 897–918.

[27] E. Castro, R. D. Hjelm, S. M. Plis, L. Dinh, J. A. Turner, V. D. Calhoun, Deep independence network analysis of structural brain imaging: application to schizophrenia, IEEE transactions on medical imaging 35 (7) (2016) 1729–1740.

[28] J. Ventura, G. S. Hellemann, A. D. Thames, V. Koellner, K. H. Nuechterlein, Symptoms as mediators of the relationship between neurocognition and functional outcome in schizophrenia: a meta-analysis, Schizophrenia research 113 (2-3) (2009) 189–199.

[29] E. Goceri, Diagnosis of Alzheimer's disease with sobolev gradient-based optimization and 3d convolutional neural network, International journal for numerical methods in biomedical engineering 35 (7) (2019) e3225.

[30] M. Hu, X. Qian, S. Liu, A. J. Koh, K. Sim, X. Jiang, C. Guan, J. H. Zhou, Structural and diffusion mri based schizophrenia classification using 2d pretrained and 3d naive convolutional neural networks, Schizophrenia research (2021).

[31] K. Oh, W. Kim, G. Shen, Y. Piao, N.-I. Kang, I.-S. Oh, Y. C. Chung, Classification of schizophrenia and normal controls using 3d convolutional neural network and outcome visualization, Schizophrenia research 212 (2019) 186–195.

[32] M. Hu, K. Sim, J. H. Zhou, X. Jiang, C. Guan, Brain MRI-based 3d convolutional neural networks for classification of schizophrenia and controls, in: 2020 42nd Annual International Conference of the IEEE Engineering in Medicine & Biology Society (EMBC), IEEE, 2020, pp. 1742–1745.

[33]    M. N. I. Qureshi, J. Oh, B. Lee, 3d-cnn based discrimination of schizophrenia using resting-state fMRI, Artificial intelligence in medicine 98 (2019) 10–17.

[34]    M. Hu, X. Qian, S. Liu, A. J. Koh, K. Sim, X. Jiang, C. Guan, J. H. Zhou, Structural and diffusion MRI based schizophrenia classification using 2d pretrained and 3d naive convolutional neural networks, Schizophrenia research 243 (2022) 330–341.

[35]    M. Zareapoor, P. Shamsolmoali, et al., Application of credit card fraud detection: Based on bagging ensemble classifier, Procedia computer science 48 (2015) (2015) 679–685.